# Feature Extraction and Classification from Planetary Science Datasets enabled by Machine Learning

Conor A. Nixon
Planetary Systems Laboratory
NASA Goddard Space Flight Center
8800 Greenbelt Road
Greenbelt, MD 20771
301-286-6757
conor.a.nixon@nasa.gov

Zachary Yahn
University of Virginia
Computer Science Department
351 McCormick Road
Charlottesville, VA 22904
770-833-9640
zry2yz@virginia.edu

Ethan Duncan
School of Information
University of California, Berkeley
102 South Hall # 4600
Berkeley, CA 94720
(510) 642-1464
ejduncan@berkeley.edu

Ian Neidel
Yale University
Timothy Dwight College
345 Temple St
New Haven, CT 06511-8238
ian.neidel@yale.edu

Alyssa C. Mills
Baylor University
Geosciences Department
One Bear Place #97354
Waco, TX 76798-7354
(254) 710-2361
Alyssa_Mills1@baylor.edu

Benoît Seignovert
Nantes Université
Laboratoire De Planétologie Et Géosciences
2 Chem. de la Houssinière Bâtiment 4, 44300 Nantes, France
+33 2 51 12 52 67
research@seignovert.fr

Andrew Larson
Business Intelligence Division
Inntopia,
782 Mountain Rd, Stowe, VT 05672
(802) 253-2905
drewlarsen27@gmail.com

Kathryn Gansler
University of Maryland
Department of Geology
8000 Regents Dr #237,
College Park, MD 20742
(301) 405-4082
ganslerk@terpmail.umd.edu

Charles Liles
NASA Langley Research Center
Center Operations Directorate
1 NASA Drive
Hampton, Virginia 23666
757-864-3157
charles.a.liles@nasa.gov

Catherine C. Walker
Department of Applied Ocean Physics and Engineering
Woods Hole Oceanographic Institution
98 Water Street
Woods Hole, Massachusetts 02543
(508) 289-3848
cwalker@whoi.edu

Douglas M. Trent
SAIC | East-2
NASA Langley Research Center
OCIO Information, Data & Analytics Services
1 NASA Drive
Hampton, Virginia 23666
(941) 621-4535
douglas.m.trent@nasa.gov

John Santerre
School of Information
University of California, Berkeley
102 South Hall # 4600
Berkeley, CA 94720
(510) 642-1464
john.santerre@ischool.berkeley.edu

*Abstract—* In this paper we present two examples of recent investigations that we have undertaken, applying Machine Learning (ML) neural networks (NN) to image datasets from outer planet missions to achieve feature recognition. Our first investigation was to recognize ice floes (also known as rafts, plates, polygons) in the chaos regions of fractured ice on Europa. We used a transfer learning approach, adding and training new layers to an industry-standard Mask R-CNN (Region-based Convolutional Neural Network) to recognize labeled floes in a training dataset. Subsequently, the updated model was tested against a new dataset, achieving 30% precision. In a different application, we applied the Mask R-CNN to recognize clouds on Titan, again through updated training followed by testing against new data, with a precision of 95% over 369 images. We evaluate the relative successes of our techniques and suggest how training and recognition could be further improved. The new approaches we have used for planetary datasets can further be applied to similar recognition tasks on other planets, including Earth. For imagery of outer planets in particular, the technique holds the possibility of greatly reducing the volume of returned data, via onboard identification of the most interesting image subsets, or by returning only differential data (images where changes have occurred) greatly enhancing the information content of the final data stream.

TABLE OF CONTENTS

## 1. INTRODUCTION

Planetary Science may be defined as the study of Solar System bodies, usually excluding the Earth and Sun, which have their own specialist fields (Earth Sciences and Heliophysics, respectively). An exception to the Earth and Sun exclusions is for the cases where their interactions with the other planets, moons and small bodies is under investigation, e.g., solar wind causing aurora on Jupiter, or study of gravitational evolution of the Earth-Moon system.

Until recently, returned image datasets from planetary science missions have been modest, due to the relatively small numbers of missions, and limitations in downlink bandwidth through the Deep Space Network (DSN) [2-4]. Their sparsity has allowed for multiple generations of researchers to conduct geophysical analysis by manual examination of images to classify features such as craters, volcanoes and rifts. The future of planetary science bodes to be different, with increasing numbers of missions from larger numbers of space agencies participating in interplanetary research voyages, and vastly larger amounts of data being returned for analysis. As with Earth remote sensing, this changing exploration landscape requires a paradigm shift in how planetary science is conducted, and most especially making use of advanced computational tools to leverage the expertise of human researchers, and to make analysis of very large datasets tractable.

In this paper we discuss recent progress by our group on the application of Machine Learning to problems in planetary science, in particular to recognition of features on images. We discuss two applications from different datasets, with differing target bodies and feature types: (i) recognition of ice floes in Europa Chaos from Galileo image data [5], and (ii) recognition of clouds on Titan from Cassini image data [6].

Our present work builds on previous work at NASA Langley Research Center (LaRC) on machine learning applications to imaging and remote sensing data. This includes an application to detect open parking spaces at NASA centers [7], and one for the detection of above anvil cirrus plumes (tops of thunderstorms) using GOES-16 visible and infrared satellite image data [8, 9]. More recently ML has been applied to monitor the health of coral reefs using CALIPSO (Cloud-Aerosol LIDAR and Infrared Pathfinder Satellite Observations) satellite data [10].

In Section 2 we give a brief introduction to artificial intelligence techniques in image recognition, followed in Section 3 by a discussion of previous AI applications in planetary science. In sections 4 and 5 we discuss our new results for Europa ice floes and Titan clouds respectively. In Section 6 we draw conclusions and suggest areas of further work.

## 2. BRIEF PRIMER: IMAGE RECOGNITION TECHNIQUES IN MACHINE LEARNING

*Machine Learning* (ML) is usually categorized as a sub-field of the broader discipline of *Artificial Intelligence* (AI) in computer science. The goal of AI is to create 'human-like' intelligence, typically through some type of computer program that may or may not have any prior knowledge base - i.e., it could be an *ab initio* creation of the programmer based on logic, statistics etc. Many types of AI are based on *Neural Networks* (NN), which are networks of nodes linked by weights that are loosely based on the brain architecture of living organisms, i.e. networks of neurons (Fig. 1). A Neural Network is said to be *fully connected* if all the neurons in each layer are connected to the previous layer.

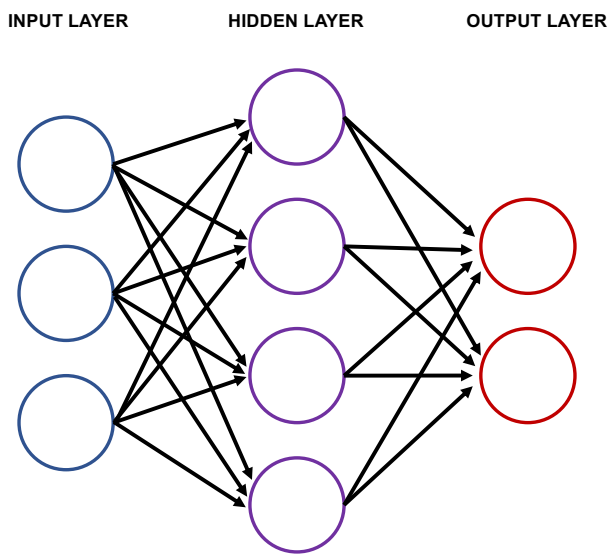

**Figure 1: Architecture of a simple 'feed forward' neural network (NN), showing the key properties of input, hidden and output layers. A NN may have multiple hidden layers.**

ML in particular is a knowledge or learning-based type of AI: a computer program is fed a set of data (images, sounds, handwriting, chess positions *etc*) as inputs, from which it is asked to make predictions or identifications such as faces, words, moves *etc* as outputs. The program is then given information about the correctness of its outputs and allowed to store this knowledge in some form to 'learn' from its past efforts and to improve performance (Fig. 2). Many ML programs use a NN architecture to store knowledge as weights between the input and output layers.

In recent years, a particular type of AI has gained prominence: *Deep Learning* (DL) based on *Neural Networks* (NN). The term Deep Learning refers to a neural network with multiple hidden layers. One particular type of NN is particularly relevant, the Convolutional Neural Network which is designed to extract higher-order features (such as faces, cats, bicycles etc in image data) from the input data (Fig. 3). A CNN has an architecture specifically inspired by an animal visual cortex. The *convolutional* nature of a CNN arises from the fact that a square window (or kernel) of fixed size, e.g. 3×3 pixels, is moved across the image, and at each position multiple filters are tested for degree of match to the data. The weights in each filter determine what feature can be detected, and the weights specifying the filters are learned by training the network against labeled data.

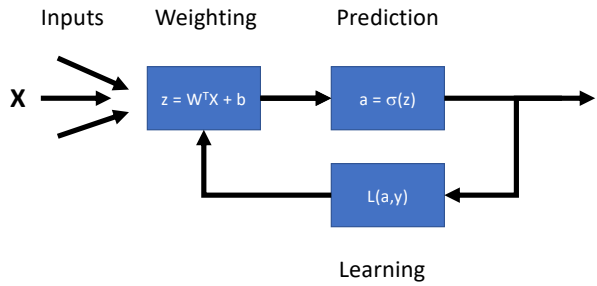

Figure 2: Architecture of a single neuron in a Neural Network. Inputs (X) are weighted (W) and bias (b) is added, before passing the output (z) to an activation function (σ) which produces a prediction (a). The result is also compared to prior knowledge (y) in an evaluation step, where a Loss Function (L) is computed that determines the quality of the result. When L is large, weights are adjusted significantly for the next iteration. As learning improves, L diminishes and the weights approach optimal values for a given dataset.

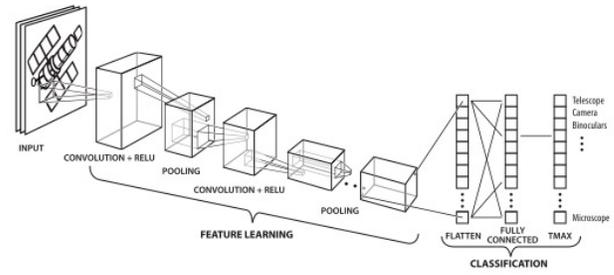

Figure 3: Architecture of a Convolutional Neural Network (CNN), showing key features: convolution layers and pooling layers for feature learning, and final classification layers. Image NASA/B. Griswold.

CNNs have been used in image recognition tasks to *classify* images and to recognize objects in the images [11]. Classification is the more straightforward task. At the most basic level it is binary: e.g., does the image contain a face (yes/no)? This often requires *object detection* to recognize objects within the image (e.g., a face, a traffic cone, a car) which can be localized with a *bounding box* - ie., the coordinates of a rectangular perimeter enclosing the specified feature ('traffic cone'). More difficult is *instance segmentation*: in this task, every pixel of the image is classified separately as belonging to the desired feature or not. So, the pixels of a face would be labeled as 'face' pixels while those of a car would be 'car' pixels.

An early significant success of a CNN was LeNet-5 [12], which was designed for handwriting recognition. The commencement of the ImageNet Large Scale Visual Recognition Challenge (ILSVRC) in 2010 led to rapid developments in NN. In the inaugural year, the winning AI had a 28% error rate on the standardized visual recognition tasks. By 2012, the AlexNet program achieved a dramatic drop in error rate to 16.4%, and by 2015 the ResNet had taken the winning error rate to 3.6%, outperforming a benchmark human score (5%).

A major advance in rapid image segmentation in the past five years has been the development of the Mask R-CNN model [13-15], a modification of an earlier model (Fast/Faster R-CNN). The traditional R-CNN (Region-based CNN) is designed to pay attention only to smaller sub-areas of the image (called *regions of interest*, RoI) and to classify each individual RoI using a CNN. Mask R-CNN adds a parallel branch of code for instance segmentation while the existing R-CNN is simultaneously producing object recognition with bounding box. The key advance here was the parallel approach to object recognition and instance segmentation, as shown in Fig. 4.

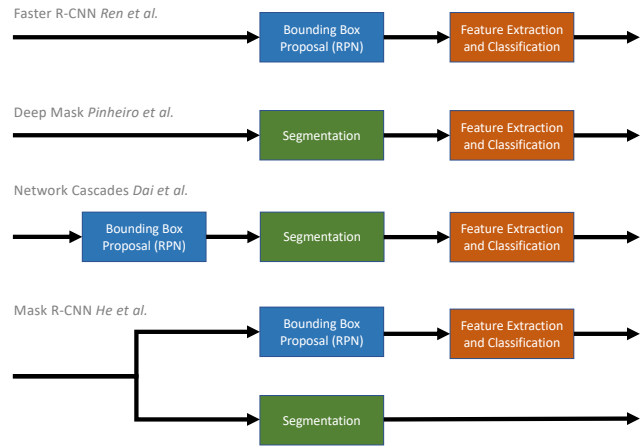

Figure 4: Various proposed ML architectures for adding segmentation task to basic classification task as implemented in *Faster R-CNN*. Note that Mask R-CNN uniquely takes a parallel approach.

We conclude our discussion of image recognition techniques by defining some standard success criteria that will be used throughout this paper:

$$Accuracy = \frac{TP + TN}{TP + TN + FP + FN}$$

$$Precision = \frac{TP}{TP + FP}$$

$$Recall = \frac{TP}{TP + FN}$$

Where the terms refer to:

TP = True positive identifications
TN = True negative identifications
FP = False positive identifications

FN = False negative identifications

## 3. PREVIOUS IMAGE RECOGNITION APPLICATIONS IN PLANETARY SCIENCE

Machine Learning (ML) applications in Planetary Science are already diverse and may be divided into the broad categories of engineering applications, and scientific applications. Engineering applications include topics such as spacecraft autonomy and fault recovery, while scientific applications are largely focused on analysis of returned data. In the future, these somewhat distinct applications are expected to converge in the quest for more efficient instrument operations, with on-board data analysis being fed back into intelligent instrument tasking, scene selection and selective data downlink [16]. In this paper the applications we discuss are exclusively feature recognition of returned images, however we note that in future this type of application may merge with onboard instrument operations. For now, we will restrict our literature review to previous applications of image recognition to keep in line with the scope of the paper.

*Crater detection:* The automated detection of craters - one of the most easily recognizable types of surface feature on solid surface planets and moons – was identified early as an obvious application for AI techniques. However even this 'simplest' problem poses challenges due to a wide range of crater sizes, morphologies (e.g. central peak vs no peak) and ages, which may lead to degradation of the rim or over-printing with younger craters.

The use of neural networks for crater detection was attempted as far back as 2005 [17], when four different approaches to search for craters in Viking image data of Mars were compared. The most successful of these, the Support Vector Machine (SVM) approach outperformed other approaches such as Feed Forward Neural Network (FFNN) and Continuously Scalable Template Model (CSTM). However the NN only detected 60% of true craters, and so was deemed at the time to be more useful as pre-filter for a human, rather than successful enough to be used as a fully autonomous technique. Nevertheless, this early work set the stage for later work with improved fidelity.

Rapid developments in AI since then, including the development of CNNs, promoted a renewed interest in automated crater counting beginning around 2015 (see recent review in [18]).[1] Initially, AI techniques were focused on image classification, such as the work of Emami et al. [19] who used a two-step approach consisting of a non-AI quantifier to identify candidate crater regions, followed by a NN to provide the final classification. A similar approach was taken by [20] to identify other features on Mars images: volcanic rootless cones and transverse aeolian ridges.

Martian crater recognition was later extended to segmentation using a U-Net based approach to identify craters in the size range 2-32 km in infrared data from the THEMIS (Thermal Emission Imaging System) instrument on the Mars Odyssey spacecraft [21], achieving a precision of ~85% and recall of ~67%.

In another approach, Lagain et al. [22] trained a NN to detect small craters on Mars with the goal of improving crater aging. And more recently the Inception-V3 image classifier, pretrained on 1.2 million images from ImageNet, was used in a transfer learning application to recognize fresh craters on Mars, succeeding with scores of 0.98 (precision), 0.975 (accuracy) and 0.93 (recall).

*Beyond Craters*: In 2020, Wilhelm et al. [23] tackled the wider problem of landform recognition on Mars, with the aim of greatly speeding up geological mapping, a process which could take years for experienced geologists to produce new wide-area maps. The team trained six different NN in a supervised learning approach to using image data from the Context Camera (CTX) on the Mars Reconnaissance Orbiter (MRO) [23]. The NN was able to recognize 15 different surface feature classes on Mars - including craters but also other landform types such as dunes, cones and ridges – providing encouragement that the tedious task of geological mapping will be able to be greatly accelerated in future with the assistance of ML techniques.

*Cassini Imagery*: While we have focused in this review section on Mars, which is heavily studied due to large available datasets of high-resolution imagery, ML image recognition has been extended to other bodies in the solar system. Yang et al. (2018) applied the Extreme Learning Machine (ELM) technique of Huang et al. [24] to edge-detection of objects (moons) in Cassini images, achieving a 94% accuaracy. The same group followed up in 2020 tackling the problem of contour detection on disk-resolved objects, used to find the center. In this work [25] they showed that Hierarchical-ELM outperformed SVM, ELM and D-CNN. Finally in 2021 AlDabbas and Gal showed that Deep CNN can be applied to Cassini images provide diverse image classification.

## 4. PROBLEM 1: DETECTING ICE PLATES IN EUROPA'S CHAOS TERRAINS

*Scientific Background:* Europa is one of the four large moons of Jupiter known as the 'Galilean Satellites', in honor of Galileo Galilei who first saw them through his telescope at the University of Padua in 1610. Europa is the smallest of the four at 3122 km diameter, and orbits Jupiter in 85.2 hours (3.55 Earth days). Locked in an orbital resonance with Io and the other large moons, Europa is prevented from fully circularizing its orbit, and therefore experiences varying tidal forces around its orbit. The variation in distance from Jupiter causes tidal heating of the interior, which leads to surface fracturing (e.g. 'cycloids') [26-28] and resurfacing, keeping the surface relatively young (20-180 Myr) [29, 30]. Europa's interior water ocean [31, 32], detected through its magnetic field signature by the Galileo spacecraft [33-35], is thought to lie 15-25 km below the surface and extend to a depth of 60-150 km relative to the surface.

---

[1] We note that the authors of this review were able to list all 13 papers then published that applied AI techniques to crater recognition in a single table – a situation unlikely to hold true for long.

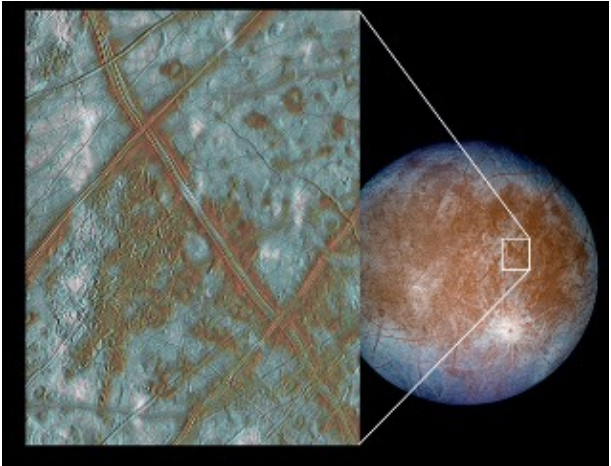

Figure 5: Europa, Conamara Chaos Region as seen by Galileo on 09/07/96, 12/96, and 02/97. Image PIA03002 Credit: NASA/JPL-Caltech

In some parts of the surface there appear to be areas of ice floes, perhaps recently mobile, that are now refrozen into a surface matrix. These 'chaos regions' (e.g. Conamara Chaos, Fig. 5) [36-41] are areas of intense scientific interest and focus, since they may be the areas where interior water has most recently been frozen onto the surface, and hence the best places to search for signatures of sub-surface ocean life without having to drill through the crust.

*Objective:* The objective of our first project was to provide automated recognition of ice blocks on Europa chaos regions. These appear as jagged polygonal features, with the larger blocks often displaying a distinctive imprinted lineation, apparently relics from the pre-disruption surface.

*Feature Definition*: Spaun et al. (1998) [42] in their detailed study of early Galileo images arrived at a working classification scheme for different features with the chaos. The scheme divided the features into two broad types: the large 'textured polygons', and 'matrix' consisting of four sub-types of successively smaller sizes, listed Table 1. Everything else surrounding the features was 'background'. Examples are shown in Fig. 6.

Rigorously determining the various feature types was not trivial for human scientists, and would be very challenging for a computer algorithm. In particular, early tests with human labeling of features showed us that different viewers would impose different cut-offs between the different block types. We therefore decided to focus only on the largest blocks ('polygons') and to formalize the definition as much as possible, adding additional criteria for identification to the simpler definition of Spaun et al. [42]. In our definition a polygon must have:

1) 'recognizable linear textures' (Spaun et al. 1998)

Table 1: Europa chaos feature types: our definition adapted from Spaun et al. (1998).

| Category | Type | Features |
|---|---|---|
| **Polygon** | Polygon | Large pieces with well-defined surface lineation, often in multiple directions. |
| **Matrix** | Micro-polygons | Medium sized pieces with poorly-defined lines. |
| **Matrix** | Angular blocks | Small pieces where the width in one dimension has approaches the height above the background, implied by shadowing. |
| **Matrix** | Peaks | Even smaller pieces where both length and width appear similar to height. |
| **Matrix** | Hummocky | Smallest distinguishable floating fragments. |
| **Background** | Background | Non-feature material within the chaos. |

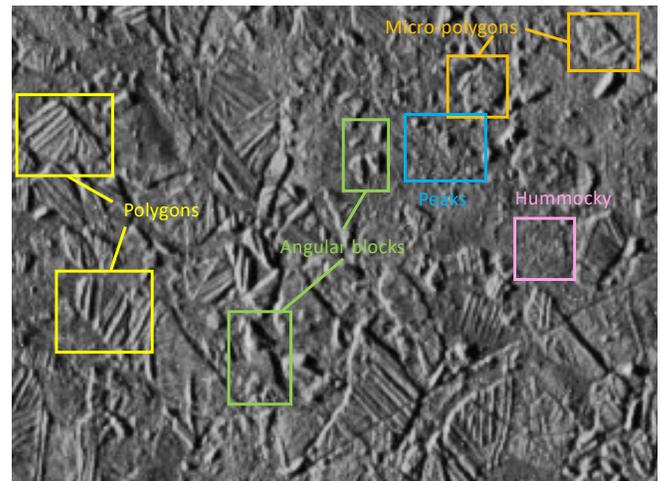

Figure 6: Examples of chaos feature types under our definition, adapted from that of Spaun et al. (1998). Image credit: NASA/JPL

(e.g., ridges, grooves, bands) that are 'usually sharp and crisp'. But also:

2) A well-defined perimeter.

3) Evidence of elevation above the surrounding matrix/mélange, which leads to shadowing under low angle sun conditions.

*Labeling*: The USGS Astrogeology Science Center provided 92 photogrammetrically-controlled global image mosaics of Europa. For our training dataset, we identified chaos regions in the Galileo SSI RegMaps ('Regional Maps') that were ≥ 50 km in length along the chaos' long axis (see Fig. 7 for full planet map). An example of such a chaos region is shown in the high-resolution image of the Conamara Chaos is shown in Fig. 8. We used the RegMaps because the resolution is roughly the same through the images at about 250 m/pixel. We also chose the RegMaps because we can look at the differences between the Trailing hemisphere and Leading hemisphere of Europa and also at latitudinal differences.

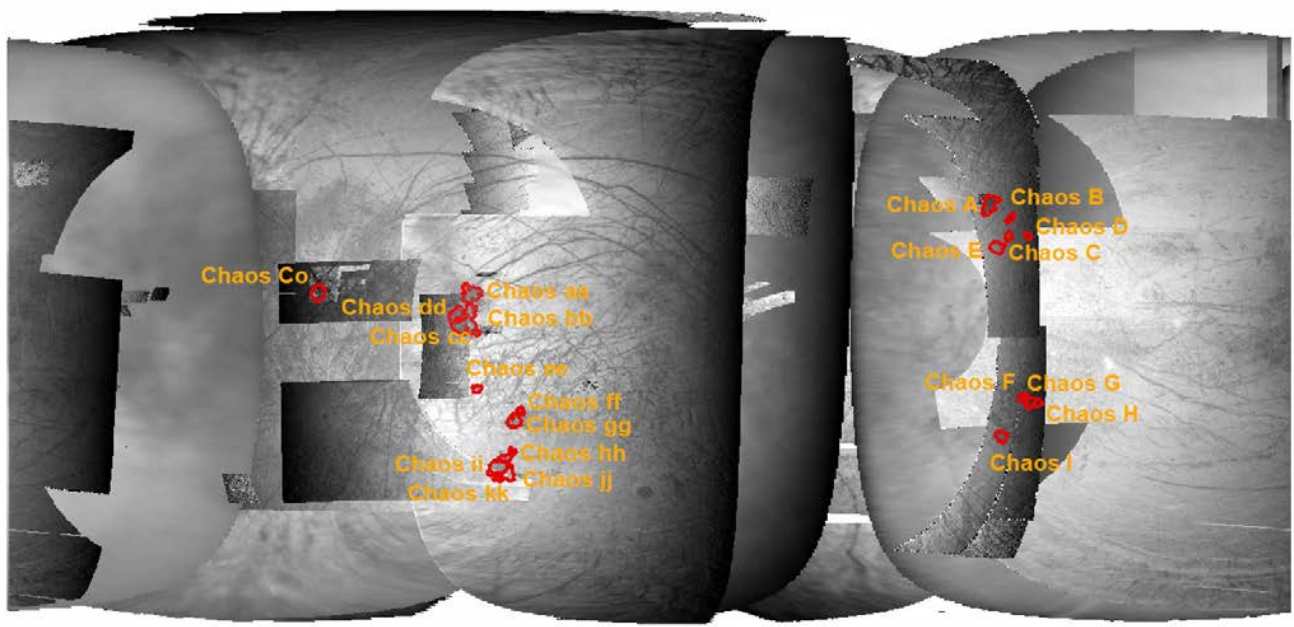

**Figure 7: Europa global mosaic from Galileo SSI data, with chaos regions identified. Names are arbitrary but our naming convention follows that of Leonard et al., 2022 [1].**

Armed with a robust definition of a polygon, we proceeded to identify individual chaos blocks (green in Fig. 8) that were ≥ 2 km in diameter, based on the resolution limitation for the given images, and met our three criteria given in the previous section. We labeled polygons in N chaos regions by vertex identification in Europa maps in ArcGIS. Within our labeling process, we also add information into the Attribute Table for each polygon, or chaos block, about each block's physical characteristics, such as area, as well as location and morphology type (plate vs knobby as per the definitions of [1]).

A histogram of identified block sizes is shown in Fig. 9. Further geophysical investigation of the size distribution will be given in a later paper.

*Machine Learning*: Having arrived at a fully labeled dataset, the next step was to match the data to the input layer of our machine learning model (Mask R-CNN). First some preprocessing of in the input data was necessary so that it can be interpretable for the Mask R-CNN model. This included exporting the .tiff image files and .shp label files from ArcMap, part of the ArcGIS suite, and converting the images into 750 dpi resolution images. The initial batch of data was restricted to the 17ESNERTRM01_GalileoSSI_Equi-cog.tif image file which had a resolution of 210 meters per pixel.

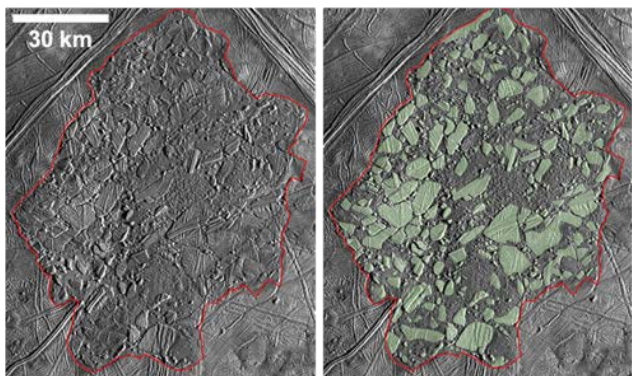

**Figure 8: Conamara Chaos image (left) and human labeled view (right).**

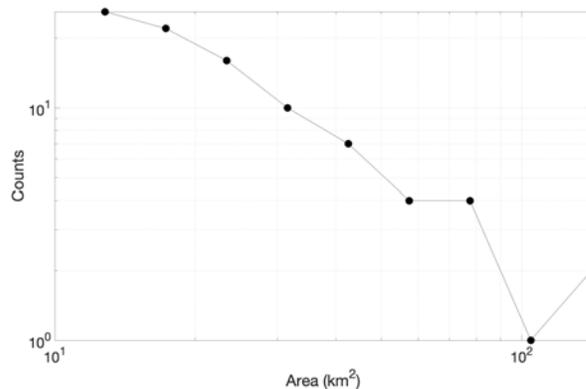

**Figure 9: Chaos block size distribution from human labeling.**

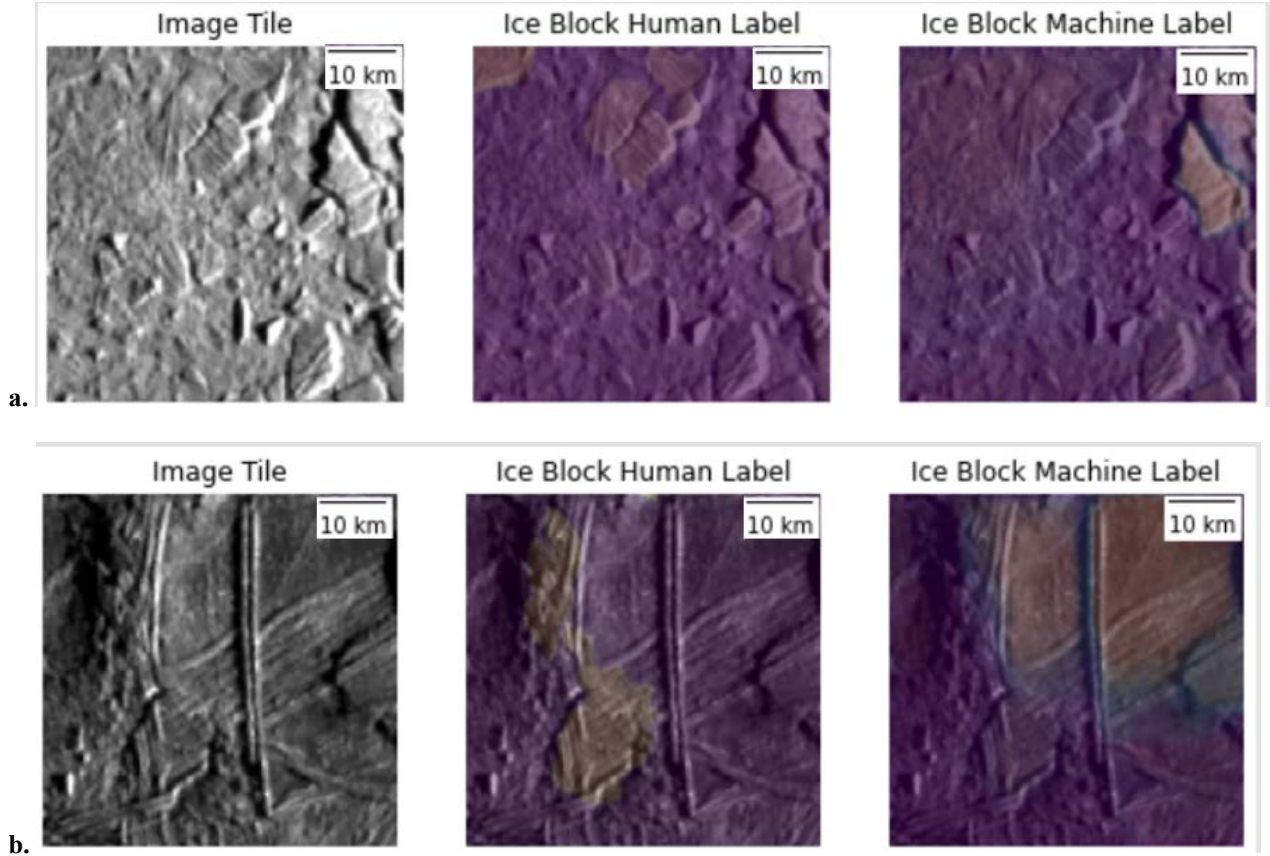

**Figure 10: (a)** Correct classification of ice blocks in chaos cc region on Europa. **(b)** Incorrect classification of ice blocks in chaos cc region on Europa

The next step was to load the images and labels onto the Google Cloud Platform for increased computing power that was later utilized for training and inference of the machine learning model. The images were then meticulously cropped to avoid included jagged image edges and image gaps that are a remote sensing artifact. We randomly broke up the image into ~50×50 km image tiles that only included one or more ice blocks. Then we split the data into 80% training, 10% testing, and 10% validation sets, which is standard industry practice.

Finally, there was an additional instance segmentation step required for the preprocessing this dataset. Due to the status of the labels pulled from the ArcMap platform, all the blocks were labeled with the same color. For the machine learning algorithm to identify the ice blocks as different entities, they need to be transformed into a different color. After this process is done programmatically, the image and label tiles are ready to be process in the machine learning algorithm.

Learning proceeded for the Mask R-CNN through the PyTorch architecture developed on the Google Cloud Platform. Training involved batching the augmented training dataset into batches of 4 images with a binary cross-entropy loss function. Loss converged after 20 epochs. The model infers a label of "ice block" in yellow and "background" purple for each pixel of a given image (Fig. 10).

*Results:* We evaluate the performance of the model based on two metrics:

1) How well does the model identify that an ice block within a given bounding box (binary classification)?
2) Given a positive classification for the existence of an ice block, how well does the model classify each pixel within the bounding box (instance segmentation)?

The first question can be answered by using accuracy, precision, recall which are defined as follows:

$$Accuracy = \frac{TP + TN}{TP + TN + FP + FN} = 0.75$$

$$Precision = \frac{TP}{TP + FP} = 0.33$$

$$Recall = \frac{TP}{TP + FN} = 0.30$$

An accuracy of 0.75 indicates that the model correctly classified 75% of all pixels in the test set, regardless of their true label. The precision score indicates that, of all the positive classifications made by the model, 33% were correct. Similarly, the recall score indicates that, of all the true positive labels in the test set, the model correctly identified 30% of them. In short, the model missed smaller ice blocks, often mistaking an entire image as an ice block. When strictly looking at large ice blocks, the model has a much higher accuracy. In future iterations, improvements include potentially combining previous iterations of the ResNet50 model employed that was able to strongly identify smaller ice blocks.

The second question can be answered by assessing how well the model classifies each pixel with an intersection over union (IoU) score which is defined as follows:

$$IoU = \frac{True\ Label\ \cap\ Model\ Label}{True\ Label\ \cup\ Model\ Label}$$

This score is produced for each image, therefore an average of all images IoU is used. The average IoU is again 0.33. This result is likely due to the many similarities of shadows, lines and coloration that ice blocks share with the hummocky background found within Europan Chaos regions.

## 5. PROBLEM 2: MAPPING TITAN CLOUDS

*Scientific Background*: Titan is the largest moon of Saturn, the second largest moon in the Solar System, and with a diameter of 5150 km is larger than the planet Mercury (4880 km) [43] – albeit much less dense. The most remarkable fact about Titan however is that it is the only moon in the Solar System to have a dense atmosphere. Its atmosphere is largely composed of nitrogen gas ($N_2$, ~95%) but with a substantial amount of methane ($CH_4$ ~5%) [44]. These two gases react chemically once activated by sunlight, created a dense mixture of complex organic chemicals [45], and a hazy orange smog that obscures sight of the surface at visible wavelengths [46, 47] (Fig. 11).

Titan's surface temperature (~95 K) [48-50] is close to the triple point of methane, but as air rises it cools adiabatically, so that around 15-30 km altitude methane condenses, forms clouds and rains out onto the surface [51, 52], where it re-evaporates back into the atmosphere [53]. Methane on Titan therefore plays a similar role to water on Earth as a condensable, precipitable substance that exists in multiple phases. The origin and persistence of methane on Titan remains a puzzle [54], since it is slowly destroyed by sunlight, and none should be remaining at the present day, unless somehow replenished. The questions of Titan's atmospheric origin are beyond the scope of this paper, however the reader may take away the message that improving our understanding of Titan's methane cycle is of fundamental importance to understanding Titan and the wider Solar System.

Although Titan's atmosphere is opaque at visible wavelengths, at some near-infrared wavelengths the atmosphere becomes transparent and the surface can be glimpsed. In addition, methane clouds may be seen (Fig. 12).

*Objective*: The objective of the second problem we tackled was to recognize and map the methane clouds in Titan's lower atmosphere from Cassini images. This was to prove a difficult problem in image segmentation, since the edges of clouds are by nature ill-defined.

By assigning each pixel of a given image a label of 'cloud' or 'no cloud,' an instance segmentation model would provide the location of cloud formations within the image. This information might then be used to calculate cloud areas, centroids, and other metrics.

*Dataset:* From the complete Cassini dataset we selected a subset of 798 images, roughly half containing examples of cloud formations. While Cassini provided hundreds of thousands of images, we chose this small sample to capture a variety of structures without requiring excessive pre-processing and labeling. The dataset contains examples of all common cloud formations found in the complete database. Several instances of images without clouds were selected for their atmospheric features or artifacts that resembled clouds to account for potential false positives.

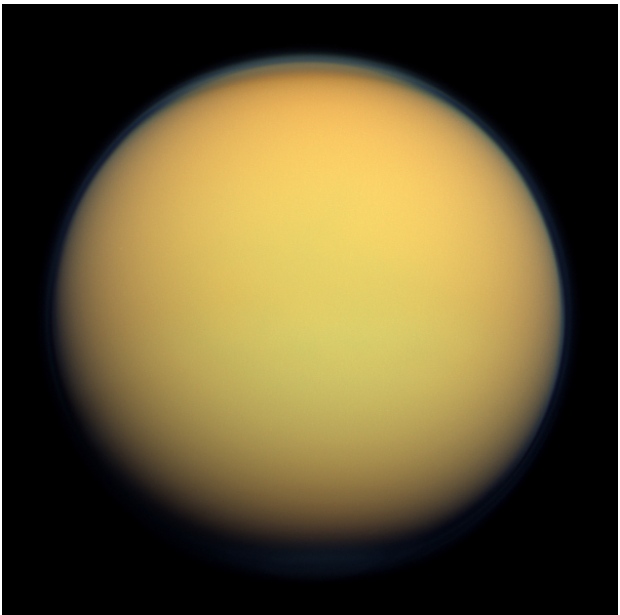

**Figure 11: Titan in natural color image by Cassini's Imaging Science Sub-system (ISS) Jan. 30, 2012 at a distance of 191,000 km. PIA14602. Image Credit JPL/NASA/Space Science Institute.**

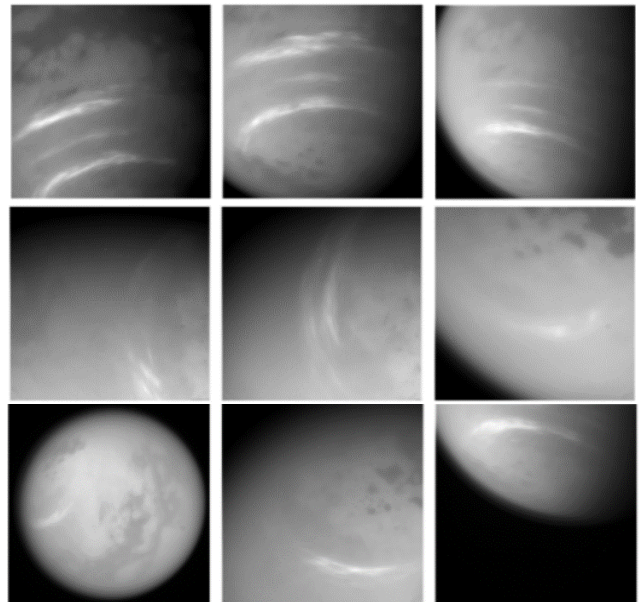

**Figure 12: Examples of clouds on Titan. Images captured by Cassini Imaging Science Subsystem (ISS) CB3 filter (938 nm) between 2014 and 2017 at a distance of $1\text{-}5\times10^5$ km.**

This dataset was then further divided into training and testing sets. The training set contains 429 images, and the testing set contains 369. Each image was hand-labeled using LabelMe software, producing a polygon for each cloud formation. The pixels enclosed in these polygons were assigned the "cloud" label, and all others were assigned the "no cloud" label.

*Machine Learning:* The model was constructed in PyTorch with code running on a Google Cloud Platform image. Training the Mask R-CNN involved batching the augmented training dataset into batches of 16 images with a binary cross-entropy loss function. Loss converged after 30 epochs. The model infers a label of "cloud" or "no cloud" for each pixel of a given image. We assign each positive cloud label a distinct color to view these pixels, producing a mask (Fig. 13)

*Results:* We evaluate the performance of the model based on two metrics:

3) How well does the model identify that a cloud formation exists within a given image (binary classification)?
4) Given a positive classification for the existence of a cloud, how well does the model classify each pixel in the image (instance segmentation)?

To address the former, we use three statistical measures: accuracy, precision, and recall. We define and compute each as follows:

$$Accuracy = \frac{TP + TN}{TP + TN + FP + FN} = 0.83$$

$$Precision = \frac{TP}{TP + FP} = 0.95$$

$$Recall = \frac{TP}{TP + FN} = 0.75$$

An accuracy of 0.83 indicates that the model correctly classified 83% of all images in the test set, regardless of their true label. The precision score indicates that, of all the positive classifications made by the model, 95% were correct. Similarly, the recall score indicates that, of all the true positive labels in the test set, the model correctly identified 75% of them. To summarize these statistics, the model may miss a true positive example of a cloud in an image, but when it does guess that a cloud exists it is almost always correct. This leaves room for improvement of our approach in future iterations, potentially by labelling more training data.

For each of the 173 images in which the model inferred the presence of at least one cloud, we assess how well it classifies each pixel with an intersection over union (IoU) score, calculated as follows.

$$IoU = \frac{True\ Label\ \cap\ Model\ Label}{True\ Label\ \cup\ Model\ Label}$$

This score is produced for each image, so we calculate the IoU of all such images by computing the average. The average IoU is 0.77. Note that this includes nine false positive cases, which each have an IoU of zero. An IoU of 0.77 indicates that the model correctly identifies a majority of cloudy pixels across all images. Upon inspecting instances where the model showed a low IoU (for example, less than 0.5), it became clear that the model struggled to identify clear edges, often identifying the correct location of the cloud but conservatively underestimating the cloud structure. This is likely a result of the ill-defined nature of cloud edges on Titan, as previously mentioned.

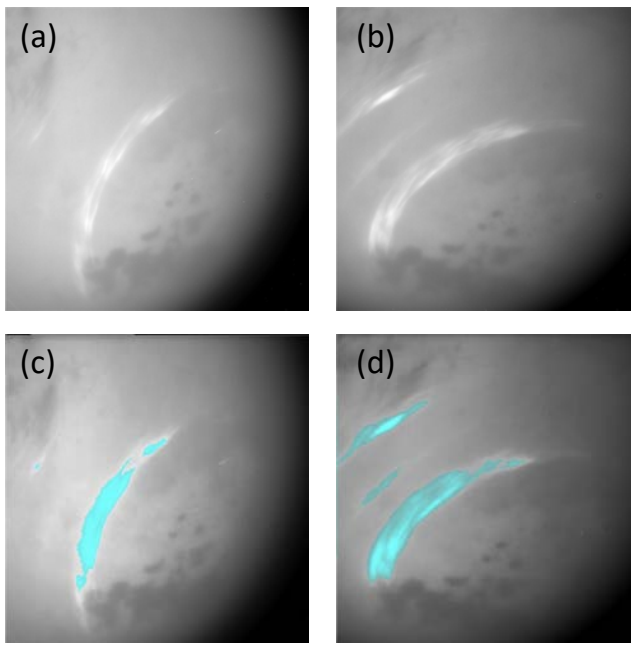

**Figure 13: Example classification of two images. Pixels labeled "cloud" are masked in blue.**

Our methods achieve comparable results to those used to identify and segment terrestrial clouds as well. Le Goff et. al use a convolutional neural network for this purpose, achieving a precision score of 81% and a recall of 75% [55]. Similarly, Li et al. fuse high-resolution satellite data from multiple sensors to achieve an average IoU of 0.9 and an accuracy of over 95% [56]. Francis et al. also demonstrate the efficacy of U-Net, another popular semantic segmentation architecture, recording a 91% accuracy [57]. Our work on Titan demonstrates similarly high scores despite the difference in image quantity and quality from terrestrial remote sensing.

## 6. SUMMARY AND CONCLUSIONS

In this paper we have described two new applications of machine learning to problems in planetary science: recognition of ice floes on Europa, and recognition of clouds on Titan. The approach we used in both cases was transfer learning used a pre-trained CNN that was subsequently minimally retrained on a training dataset before application to a test dataset. The precision obtained was 95% for Titan, showing impressive ability to replicate a human labeler. The cloud recognition approach should readily be applicable to similar problems for cloud recognition on Earth, Mars and other planets. For Europa was obtained a lower precision of 33%, which appears to be due to difficulties with consistent labeling of the desired objects. In future we believe that this can be improved by use of a more consistent set of labeled data, and we have plans to investigate crowd-sourcing as a way to achieve this. Overall, it appears that ML will have a large and increasing role to play in planetary exploration, including not just science applications but also in engineering and autonomous operations and decision making.

# APPENDICES

## A. ACRONYMS AND ABBREVIATIONS

| | |
|---|---|
| AI | Artificial Intelligence |
| BPNN | Back Propagation Neural Network |
| CALIPSO | Cloud-Aerosol Lidar and Infrared Pathfinder Satellite Observations |
| CNN | Convolutional NN |
| CSTM | Continuosly Scalable Template Model |
| CTX | Context Imager |
| DL | Deep Learning |
| DSN | Deep Space Network |
| ELM | Extreme Learning Machine |
| FFNN | Feed Forward Neural Network |
| FN | False Negative |
| FP | False Positive |
| GIS | Geographical Image System |
| GOES | Geostationary Operational Environmental Satellites |
| ILSVRC | ImageNet Large Scale Visual Recognition Challenge |
| IoU | Intersection over Union |
| KNN | $k$-Nearest Neighbor |
| ML | Machine Learning |
| MRO | Mars Reconnaissance Orbiter |
| NN | Neural Network |
| PCR | Principal Component Regression |
| PIA | Planetary Image Archive |
| R-CNN | Region-based CNN |
| SSI | Solid State Imager |
| SVM | Support Vector Machine |
| THEMIS | Thermal Emission Imaging System |
| TN | True Positive |
| TP | True Negative |
| USGS | United States Geological Survey |

## ACKNOWLEDGEMENTS

The authors thank the NASA Internship Program for their support during the completion of this work.

# BIOGRAPHY

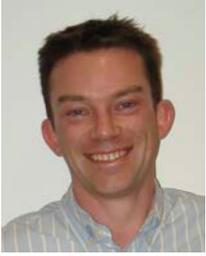
***Conor Nixon*** *is a scientist in the Planetary Systems Laboratory at NASA Goddard Space Flight Center in Greenbelt, MD. He has 20 years of experience in planetary science research, especially focusing on remote sensing of the atmospheres of the outer planets and Titan from spacecraft platforms such as Galileo and Cassini. He was the Deputy Principal Investigator of Cassini's Composite Infrared Spectrometer (CIRS) and has led mission design studies for future missions to Jupiter and Titan. He is currently engaged in instrument development for future planetary missions to the outer solar system, including next-generation infrared FTS instruments.*

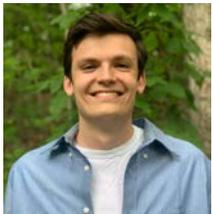
***Zach Yahn*** *is an undergraduate computer science and computer engineering major at the University of Virginia and research intern for Goddard Space Flight Center. He is broadly interested in machine learning, artificial intelligence, and their applications to real-world problems. Next year he plans to continue research in graduate school by pursuing a PhD.*

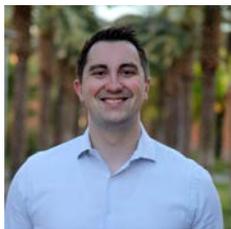
***Ethan Duncan*** *is a graduate Master of Information and Data Science student at University of California, Berkeley, and a research intern for Goddard Space Flight Center. He graduated in 2021 with dual Bachelor of Science degree in Astrophysics and Physics from Arizona State University and had an opportunity to work on a NASA STTR funding machine learning internship that focused on finding Lunar and Martian planetary features. He is interested in deep and machine learning technologies and their use in solving business and scientific problems. He plans on completing his graduate program and finding work in industry.*

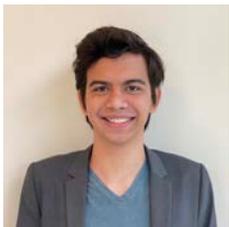
***Ian Neidel*** *is an undergraduate double majoring in computer science and global affairs at Yale University and a research assistant for Goddard Space Flight Center. He is interested in deep learning technologies and their broader societal implications. He plans on doing industry machine learning work after graduation before going back for a PhD.*

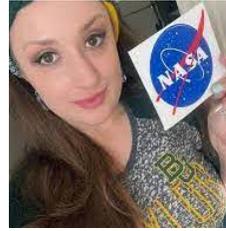
***Alyssa Mills*** *is currently a PhD student at Baylor University, studying gravity measurements of icy bodies. Formerly she was a master's student at The University of Alabama studying the magnetic field of Ganymede. She was previously an intern at NASA JPL, studying chaos on Europa. She has also been an Research Assistant at NASA Goddard Space Flight Center and Smithsonian CEPS cryovolcanic domes on Europa. She created an extensive database of cryovolcanic domes that included information on the domes' physical characteristics and geological settings. She received dual degrees in astronomy and geology (with honors) from the University of Maryland.*

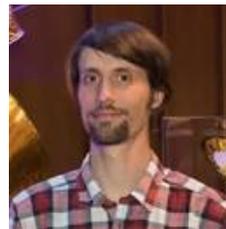
***Benoit Seignovert*** *is a planetary data scientist at the Laboratory of Planetary and Geosciences in Nantes. He is currently developing a coverage analysis tool for the European Space Agency (ESA) to improve the observation planning of the JUICE mission. Previously, he worked as a postdoc researcher in the Planetary and Exoplanetary Atmospheres group at NASA/JPL. He completed a PhD in Astrophysics in 2017 at the University of Reims where I developed radiative transfer models to analyze the detached haze layers observed by Cassini-ISS instrument at high altitude in Titan's upper atmosphere.*

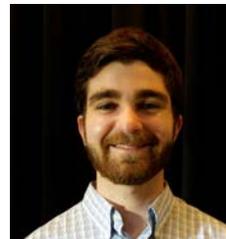
*Andrew Larsen is a data scientist at Inntopia. He graduated from the University of Vermont with Bachelor of Science degrees in Mathematics and Physics in 2018, and graduated with a Masters of Science in Data Science from Southern Methodist University in 2021. He is interested in applying machine learning and deep learning solutions to problems in both the public and private sector. Past projects include sentiment analysis, time series analysis and computer vision.*

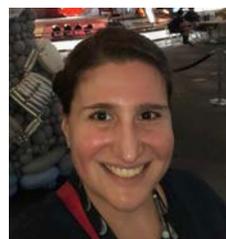
***Katie Gansler*** *is a graduate student in geology at the University of Maryland, where she studies planetary science. She graduated Magna Cum Laude from the University of Pennsylvania with a degree in political science in 2014. Several years later, she received a second bachelor's degree in astronomy from the University of Maryland. She has had four internships at NASA, one at Headquarters, one at Glenn Research Center, and the other two at the Goddard Space Flight Center.*

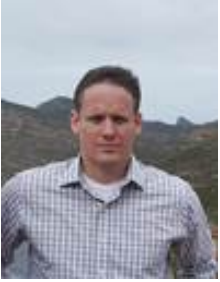***Charles Liles*** *is a Data Systems Aerospace Technologist at NASA Langley's Center Operations Directorate. He has supported the development of applied artificial intelligence and machine learning models as well as cloud technologies at NASA for 9 years. He currently supports the development of smart infrastructure technologies and the integration of enterprise data systems and analytics tools. He also serves as GIS Team Lead for Langley Research Center.*

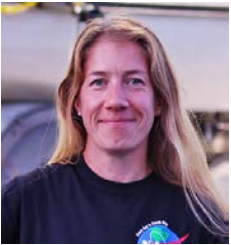***Catherine Walker*** *is a glaciologist and planetary scientist at Woods Hole Oceanographic Institution in Cape Cod, MA and a visiting research scientist in the Cryospheric Sciences Laboatory at NASA Goddard Space Flight Center in Greenbelt, MD. Her research focuses on ice dynamics, ice-ocean interactions, and fracture-fault mechanics on Earth and Ocean Worlds of the solar system. She has expertise in remote sensing methods (imagery, lidar, radar) and also uses autonomous vehicles to explorer Earth's polar oceans.*

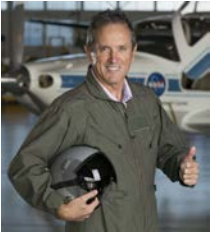***Douglas Trent*** *is a Senior Data Scientist in the Office of the CIO at NASA's Langley Research Center, leading cross-agency teams to accelerate adoption of AI/ML technologies in support of NASA's digital transformation strategy. Before joining NASA in 2018, he spent 26 years at IBM managing PC product lines worldwide and serving as the Business Development Executive for strategic relationships with Oracle and Microsoft. Douglas authored IBM's human-centered design strategy, coined the term "human-centric computing" and helped lead a small team to create the world's first smart phone.*

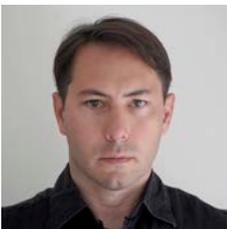***John Santerre*** *is a Lecturer at the University of California Berkeley and a Senior Adjunct Faculty at Southern Methodist University. He is also a director of Data Science at Silicon Valley Bank where he leads the enterprise AI LAB. He completed his PhD at the University of Chicago CS department, where he studied Machine Learning and its application to Computational Biology data sets at Argonne National Laboratory. He brings over 10 years of experience connecting recent research results to applied teams in academia and industry.*

17